\documentclass[a4paper,12pt]{article}
\usepackage{latexsym,indentfirst}
\newcommand{\nwc}{\newcommand}
\nwc{\be}{\begin{equation}}
\nwc{\ee}{\end{equation}}
\nwc{\bea}{\begin{eqnarray}}
\nwc{\eea}{\end{eqnarray}}
\begin{document}
\pagestyle{plain}
\title{Constraints on mass matrices due to measured property of the mixing 
matrix}
\author{S. Chaturvedi \footnote{scsp@uohyd.ernet.in}\\
\rm School of Physics\\
University of Hyderabad, Hyderabad 500 046, India\\  
Virendra Gupta \footnote{virendra@aruna.mda.cinvestav.mx}\\
\rm Departamento de Fis\'ica Aplicada, CINVESTAV-Unidad M\'erida\\
 A.P. 73 Cordemex 97310 M\'erida, Yucatan, Mexico} 
\maketitle
\begin{abstract}
 It is shown that two specific properties of the unitary matrix $V$ can be
 expressed directly in terms of the matrix elements and eigenvalues of the
 hermitian matrix $M$ which is diagonalized by $V$. These are the asymmetry 
$\Delta(V)= |V_{12}|^2- |V_{21}|^2$, of $V$ with respect to the main diagonal
 and the Jarlskog invariant $J(V)= {\rm Im}(V_{11}V_{12}^* V_{21}^* V_{22})$. 
These expressions for $\Delta(V)$ and $J(V)$ provide constraints on possible
 mass matrices from the  available data on $V$. 

\end{abstract}
\maketitle
\newpage
\section{Introduction}
Flavor mixing in both the quarks and lepton sector has been firmly established
experimentally for a long time, However, there is still no deep understanding
of the observed mixings. 

In the Standard Model, the unitary mixing matrices arise from the
diagonalization of the corresponding hermitian mass matrices. In the lepton
sector one usually works in the basis in which the charged lepton mass matrix
is diagonal, so that the neutrino flavor mixing is described by a single
unitary matrix \cite{1} which diagonalizes the neutrino mass matrix. In the
quark sector \cite{2}, in  the physical basis, the CKM-mixing matrix $V=
U^\dagger U^\prime$, where the unitary matrices $U$ and $U^\prime$ diagonalize
the up-quark and down-quark mass matrices respectively. One can also work in a
basis in which the up-quark ( down-quark ) mass matrix $M$ ( $M^\prime$ ) is
diagonal. In these bases, the mixing matrix in the quark sector ( like the
neutrino sector ) will come from a single mass matrix. Clearly, if we knew the
mass matrices fully then the corresponding mixing matrices are completely
determined. In practice, the mass matrices are guessed at, while experiment
can only determine the numerical values of the matrix elements of the mixing
matrix. Our objective is to learn something about the structure of the
underlying mass matrix $M$ from the knowledge of the mixing matrix $V$ which
diagonalizes it. Can a general property of $V$ imply a constraint on $M$ ? In
particular we show that the asymmetry $\Delta (V) $ w.r.t. the main diagonal
and, $J(V)$, the Jarlskog invariant \cite{3} which is a measure of 
CP-violation can be
directly expressed in terms of the eigenvalues and matrix elements of $M$!
These can provide simple criterion for selecting suitable mass matrices.

\section{Derivations of the formulas for $\Delta(V)$ and $J(V)$} 
 Consider a $3\times 3$ mass matrix $M$ which is diagonalized by the unitary
 matrix $V$, so that
\be
M=V~{\widehat M}~V^\dagger,
\label{1}
\ee
where ${\widehat M}= {\rm diag}(m_1,m_2,m_3)$. We can also write 
\be
M=m_1~N_1+m_2~N_2+m_3~N_3,
\label{2}
\ee
where $N_\alpha$ are the projectors of $M$. They satisfy, 
\be 
N_\alpha~N_\beta=N_\alpha~\delta_{\alpha\beta}~~~{\rm and}~~~~~
(N_\alpha)_{k\ell} = V_{k\alpha}~V_{\ell \alpha}^{*}.
\label{3}
\ee
Furthermore, in terms of $M$ and its eigenvalues, 
\be
N_\alpha=\frac{(m_\beta-M)(m_\gamma -M)}{(m_\beta-m_\alpha)
(m_\gamma -m_\alpha)},~~~\alpha\neq \beta \neq \gamma,
\label{4}
\ee
with $\alpha,~\beta,~\gamma$ taking values from $1$ to $3$. It is clear from 
Eq.$(\ref{3})$ that 
\be
|V_{k\alpha}|^2=(N_\alpha)_{kk}.
\label{5}
\ee
Through this equation each $|V_{k\alpha}|$ can be calculated in terms of the
eigenvalues\footnote {Our results require non-degenerate eigenvalues. This is
  true for the quarks.} and matrix elements of $M$.
\vskip0.3cm
\noindent
{\it (a) Formula for $\Delta(V)$}
\vskip0.3cm

The asymmetry with respect to the main diagonal of $V$ is given by 
\be
\Delta(V)\equiv |V_{12}|^2-|V_{21}|^2=|V_{23}|^2-|V_{32}|^2=
|V_{31}|^2-|V_{13}|^2.
\label{6}
\ee
The last two equations follow from the unitarity of $V$, namely, $VV^\dagger=
V^\dagger V= I$. Using Eqs.$(\ref{4},\ref{5})$, simple algebra gives, 
\be
\Delta(V)=\frac{1}{D(m)}\{\sum_{k}\left(m_k~(M^2)_{kk}-m_{k}^{2}~M_{kk}\right)
\},
\label{7}
\ee
where
\bea
D(m)&\equiv& \left |\matrix{1 & 1 & 1 \cr m_1 & m_2 & m_3 \cr m_{1}^2 
& m_{2}^2 & m_{3}^{2}}\right| \nonumber\\
&=& (m_2-m_1)(m_3-m_1)(m_3-m_2).
\label{8}
\eea  
Our result in Eq.$(\ref{7})$ tells us, given $m_i$ and $M$, whether
 $V$ will be symmetric $(\Delta(V)=0)$ or not. We note that the asymmetry 
for the CKM-matrix, $\Delta(V)$ is intriguingly small!
\vskip0.3cm
\noindent
{\it (b)  Formula for $J(V)$}
\vskip0.3cm
We use the definition
\be
J(V)= {\rm Im}(V_{11}V_{12}^* V_{21}^* V_{22}).
\label{9}
\ee 
The imaginary parts of eight other plaquettes are just $\pm J(V)$ because $V$
is unitary \cite{3}. To derive our result we note that 
\be
M_{12}M_{23}M_{13}^* = \sum_{k,\ell,n}~m_km_\ell m_n V_{1k} V_{2k}^* 
V_{2\ell} V_{3\ell}^* V_{1n}^* V_{3n} , 
\label{10}
\ee
since from Eq.$(\ref{1})$, $M_{ij}=\sum_{k}~m_k V_{ik} V_{jk}^*$. Now use 
the unitarity relation $V_{1\ell}^*V_{1n}+V_{2\ell}^*V_{2n}+ V_{3\ell}^*
V_{3n}=\delta_{\ell n}$ and take imaginary parts to obtain
\bea
{\rm Im}(M_{12}M_{23}M_{13}^*)& =&
\sum_{k,\ell}~m_km_{\ell}^2~{\rm Im}( V_{1k} V_{2k}^* 
V_{2\ell} V_{1\ell}^*)\nonumber \\ 
&-& \left[\sum_{n}~m_n(|V_{1n}|^2 +|V_{2n}|^2)\right]\cdot
\sum_{k,\ell}~m_km_{\ell}~{\rm Im}( V_{1k} V_{2k}^* 
V_{2\ell} V_{1\ell}^*)\nonumber\\
\label{11}
\eea
The imaginary parts on the RHS,  for various plaquettes of $V$, yield 
$\pm J(V)$ for various values of $k$ and $\ell$. As a result the second term
sums up to zero. One thus obtains
\be
J(V)= \frac{{\rm Im}(M_{12}M_{23}M_{13}^*)}{D(m)}.
\label{12}
\ee
This remarkable result shows that if $M_{12}M_{23}M_{13}^*$ is real for a
given M, then the Jarlskog invariant for the matrix $V$ which diagonalizes it
vanishes. Thus to obtain CP-violation, the mass matrix for up-quark 
(down- quark) must have ${\rm Im}(M_{12}M_{23}M_{13}^*)$
non-zero in a basis in which down-quark (up-quark) mass matrix is diagonal. 
Equivalently, $\Theta\equiv \theta_{12}+\theta_{23}-\theta_{13} 
\neq n\pi~(n=0,1,2,\cdots )$, here $\theta_{ij}$ is the phase of
$M_{ij}$. This is reminiscent of the fact \cite{5} that physically the
relevant phase for CP-violation in CKM-matrix is
$\Phi\equiv\phi_{12}+\phi_{23}-\phi_{13}$, where $\phi_{ij}$ is the phase of
$V_{ij}$. The reason is that $\Phi$ is invariant under re-phasing
transformations. Clearly, under re-phasing transformations $\Theta$ is also
invariant (See Eq.$(\ref{11})$). Furthermore it can be shown directly that if
any one of the three off-diagonal elements of $V$ is zero then $J(V)$
vanishes \cite{6}. Thus the appearance of the numerator in Eq.$(\ref{12})$ 
is understandable.

The use of Eqs.$(\ref{7})$ and $(\ref{12})$ for the quark sector will be
considered elsewhere.

\indent \textbf{Acknowledgments}
One of us (VG) is grateful to Dr. A. O. Bouzas for discussions. 

\newpage


\begin{thebibliography}{99}
\bibitem{6} Let $M_{ij}=m_1V_{i1}V_{j1}^* +m_2 V_{i2}V_{j2}^*
+ m_3V_{i3}V_{j3}^* =0,~i\neq j$. Multiply by $V_{i3}^*V_{j3}$ and take
imaginary parts to obtain $(m_1-m_2)J(V)=0$. Similarly, multiplying  
by $V_{i2}^*V_{j2}$ and $V_{i1}^*V_{j1}$ and  taking 
imaginary parts one  obtains $(m_1-m_3)J(V)=0$ and $(m_2-m_3)J(V)=0$. Thus
either $m_1=m_2=m_3$, that is, $M$ is trivial or $J(V)=0$.
\end{thebibliography}
\end{document}